\documentclass[%
 reprint,
superscriptaddress,
 amsmath,amssymb,
 aip,
]{revtex4-2}

\usepackage{graphicx}
\usepackage{dcolumn}
\usepackage{bm}
\usepackage{hyperref}
\usepackage{xcolor}

\usepackage[braket, qm]{qcircuit}
\usepackage[T1]{fontenc}
\begin{document}

\title{Real-time feedback protocols for optimizing fault-tolerant two-qubit gate fidelities in a silicon spin system}
\author{Nard Dumoulin Stuyck}
\email{n.dumoulin@unsw.edu.au}
\affiliation{School of Electrical Engineering and Telecommunications, The University of New South Wales, Sydney, NSW 2052, Australia}
\affiliation{Diraq, Sydney, NSW, Australia}
\author{Amanda E. Seedhouse}\email{a.seedhouse@unsw.edu.au}
\affiliation{School of Electrical Engineering and Telecommunications, The University of New South Wales, Sydney, NSW 2052, Australia}
\affiliation{Diraq, Sydney, NSW, Australia}
\author{Santiago Serrano}\affiliation{School of Electrical Engineering and Telecommunications, The University of New South Wales, Sydney, NSW 2052, Australia}
\author{Tuomo Tanttu}\affiliation{School of Electrical Engineering and Telecommunications, The University of New South Wales, Sydney, NSW 2052, Australia}
\affiliation{Diraq, Sydney, NSW, Australia}
\author{Will Gilbert}\affiliation{School of Electrical Engineering and Telecommunications, The University of New South Wales, Sydney, NSW 2052, Australia}
\affiliation{Diraq, Sydney, NSW, Australia}
\author{Jonathan Yue Huang}\affiliation{School of Electrical Engineering and Telecommunications, The University of New South Wales, Sydney, NSW 2052, Australia}
\author{Fay Hudson}\affiliation{School of Electrical Engineering and Telecommunications, The University of New South Wales, Sydney, NSW 2052, Australia}
\affiliation{Diraq, Sydney, NSW, Australia}
\author{Kohei M. Itoh}\affiliation{School of Fundamental Science and Technology, Keio University, Yokohama, Japan}
\author{Arne Laucht}\affiliation{School of Electrical Engineering and Telecommunications, The University of New South Wales, Sydney, NSW 2052, Australia}
\affiliation{Diraq, Sydney, NSW, Australia}
\author{Wee Han Lim}\affiliation{School of Electrical Engineering and Telecommunications, The University of New South Wales, Sydney, NSW 2052, Australia}
\affiliation{Diraq, Sydney, NSW, Australia}
\author{Chih Hwan Yang}\affiliation{School of Electrical Engineering and Telecommunications, The University of New South Wales, Sydney, NSW 2052, Australia}
\affiliation{Diraq, Sydney, NSW, Australia}
\author{Andre Saraiva}\affiliation{School of Electrical Engineering and Telecommunications, The University of New South Wales, Sydney, NSW 2052, Australia}
\affiliation{Diraq, Sydney, NSW, Australia}
\author{Andrew S. Dzurak}
\email{a.dzurak@unsw.edu.au}
\affiliation{School of Electrical Engineering and Telecommunications, The University of New South Wales, Sydney, NSW 2052, Australia}
\affiliation{Diraq, Sydney, NSW, Australia}

\date{\today}

\begin{abstract}
Recently, several groups have demonstrated two-qubit gate fidelities in semiconductor spin qubit systems above 99\%. Achieving this regime of fault-tolerant compatible high fidelities is nontrivial and requires exquisite stability and precise control over the different qubit parameters over an extended period of time. This can be done by efficiently calibrating qubit control parameters against different sources of micro- and macroscopic noise. Here, we present several single- and two-qubit parameter feedback protocols, optimised for and implemented in state-of-the-art fast FPGA hardware. Furthermore, we use wavelet-based analysis on the collected feedback data to gain insight into the different sources of noise in the system. Scalable feedback is an outstanding challenge and the presented implementation and analysis gives insight into the benefits and drawbacks of qubit parameter feedback, as feedback related overhead increases. This work demonstrates a pathway towards robust qubit parameter feedback and systematic noise analysis, crucial for mitigation strategies towards systematic high-fidelity qubit operation compatible with quantum error correction protocols.   

\end{abstract}

\maketitle

Recent demonstrations of silicon spin qubit systems with two-qubit gate fidelities above 99\% confirm the promise of the qubit platform for large-scale quantum computing \cite{Noiri2022AProcessors, Mills2022Two-qubit99, Madzik2022PrecisionSilicon,Tanttu2023StabilitySilicon}. However, fault-tolerant quantum computing will require consistent high-fidelity single- and two-qubit control, both over an extended period of time, and over an extended array of devices \cite{Fowler2012SurfaceComputation, Tanttu2023StabilitySilicon}. Therefore, precise qubit parameter control and a better understanding of the origin of the noise acting on the qubit system are crucial steps towards developing mitigating strategies, both in device fabrication and operation.   


Parameter feedback plays a key role in the stability and performance of qubit devices in the field of quantum computing~\cite{Vepsalainen2022ImprovingFeedback, Gilbert2023On-demandQubits, Tanttu2023StabilitySilicon,Blok2014, Philips2022}. A typical feedback protocol measures the status of a system and adjusts the system's control parameters to counteract any unwanted change. Feedback data, or control parameter correction data, is useful for several purposes. Firstly, it indicates if an experiment ran successfully, as it reflects the stability of different qubit parameters during the run time. Next, feedback can improve key characteristics of the system being studied, such as the quality factor of Rabi oscillations in the case of qubit data~\cite{Vepsalainen2022ImprovingFeedback, Gilbert2023On-demandQubits}. Finally, the feedback data contains valuable information about noise signals impacting the system and their dynamics. Understanding these signals can lead to improvements in qubit control and device fabrication, but efficiently extracting information is still an outstanding challenge \cite{Rojas2023}. 

One promising technique for advanced noise analysis is wavelet analysis as it allows to characterise and analyse the complex dynamics of quantum systems~\cite{Percival2000WaveletAnalysis,Seedhouse2023Wavelet-basedSystems}. Quantum systems typically display both time dependent and independent signals. Time dependent signals can be stationary with a constant periodicity, or non-stationary, with a periodicity that changes over time. Wavelet analysis can identify non-stationary signals providing insight on the duration, frequency and occurrence time(s) of an event, making it more powerful and better suited for qubit feedback data than a standard Fourier analysis~\cite{Almog2016DynamicNoise,Chan2018AssessmentSpectroscopy,Guo2022}. 

In this Letter, we present fast feedback protocols for several single- and two-qubit parameters for spin qubits in a silicon quantum dot system. The feedback protocols are optimised for fast FPGA based hardware and are executed in real-time. We then demonstrate how wavelet transformations on the feedback protocol data can help identify different noise characteristics in the qubit system, and give additional insight compared to the well-established Fourier transform. 

\begin{figure}
    \centering
    \includegraphics[width=0.8\columnwidth]{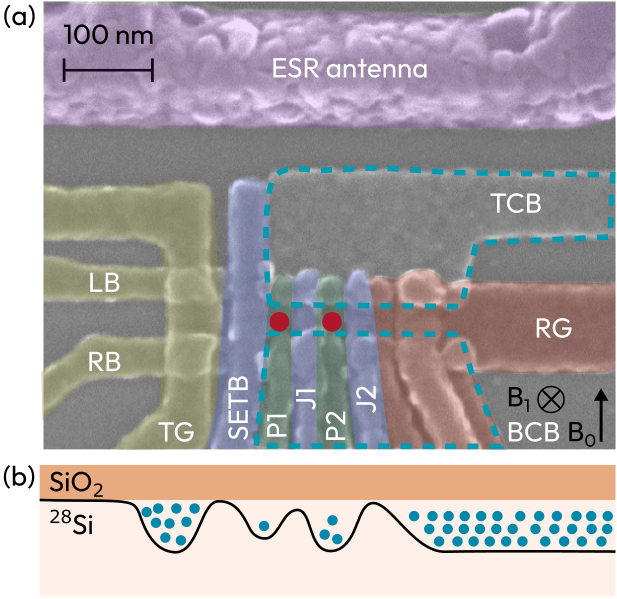}
    \caption{(a) False coloured SEM image of a device nominally identical to the one used in this letter. (b) Schematic diagram of the cross section of the device, red dots show the position of qubits 1 (Q$_1$) and 2 (Q$_2$).}
    \label{fig:experiment}
\end{figure}



Figure~\ref{fig:experiment}(a) shows a scanning electron microscope image of a device nominally identical to the one used in the experiment, accompanied by a schematic cross-section in Figure~\ref{fig:experiment}(b). The device accumulates electrons at the interface between the SiO$_2$ dielectric and isotopically enriched $^{28}$Si substrate (800 ppm) using gate electrodes fabricated in an Al/AlOx gate stack \cite{Gilbert2023On-demandQubits}. We operate the device with a fixed number of electrons in the so-called `isolation' mode \cite{Yang2020OperationKelvin}. The left and right dots are formed under plunger gates P1 and P2 containing one electron and three electrons, respectively, written as (1,3) \cite{Yang2020OperationKelvin,Seedhouse2021PauliControl}. Top and bottom confinement barrier gates, TCB and BCB, respectively, surround the dots laterally. Exchange interaction between the quantum dot sites is activated by applying a voltage to the interstitial J1 gate. An in-plane applied static magnetic field $B_0$ of 0.7 T splits the spin up and down states of the unpaired electrons, forming the two qubit states. Spin readout is performed using a Pauli spin blockade-based parity readout at the (1,3)-(0,4) charge configuration\cite{Seedhouse2021PauliControl}. We perform single-qubit gates by applying microwave (mw) pulses to an on-chip electron spin resonance (ESR) antenna. These pulses generate bursts of an alternating magnetic field $B_1$ perpendicular to $B_0$. The spin state rotates when the mw pulse frequency matches the qubit Larmor frequency \cite{Veldhorst2015ASilicon,Veldhorst2017SiliconComputer}. An arbitrary wave generator (AWG) generates gate voltage pulses for qubit initialisation and readout, and controls the IQ modulation of a microwave vector source carrier frequency. The IQ modulation controls microwave pulse time, frequency, amplitude, and phase for individual qubit manipulation. The used FPGA hardware allows for real-time parameter changes, enabling fast feedback protocols.

\begin{table*}
\caption{Overview of the different qubit parameters, their corresponding experimental control parameter, and feedback measurement protocol. See main text for individual protocol details. Q$_i$ refers to the target qubit, where $i = 1,2$. Pauli Spin Blockade (PSB) is used for parity spin readout\cite{Seedhouse2021PauliControl}. Control parameters are microwave (mw) amplitude, intermediate frequency (IF) frequency and phase, exchange J gate voltage, and charge detuning voltage.}
\label{tab:feedback}
\begin{center}
\begin{ruledtabular}
\begin{tabular}{ l l l }
   Qubit parameter & Control parameter & Feedback protocol \\ 
  \hline\\
   Rabi frequency Q$_i$ & mw amplitude Q$_i$ & \hspace{2em} \Qcircuit @C=1em @R= 0.7em { 
\lstick{\ket{\textrm{Q}_i}}  & \gate{\sqrt{\textrm{X}}^{n,n + 2}} &  \meter & \qw} \\ 
\\ \hline \\
  Larmor frequency Q$_i$ &  IF frequency Q$_i$ & \hspace{2em} \Qcircuit @C=1em @R= 0.7em {
\lstick{\ket{\textrm{Q}_i}} & \gate{\sqrt{\textrm{X}}} & \gate{t_\textrm{wait}} & \gate{\pm \sqrt{\textrm{Y}}} & \meter & \qw} \\ 
 \\ \hline \\
   Exchange level & J gate voltage & \hspace{2em} \Qcircuit @C=1em @R= 0.7em {
\lstick{\ket{\textrm{Q}_1}} & \gate{\sqrt{\textrm{X}}} &  \multigate{1}{\textrm{CZ}^n} & \gate{\pm \sqrt{\textrm{Y}}} & \multigate{1}{\textrm{PSB}} & \qw \\
\lstick{\ket{\textrm{Q}_2}} & \gate{\textrm{X}} & \ghost{\textrm{CZ}^n} & \qw & \ghost{\textrm{PSB}} & \qw } \\
 \\  \hline\\
   Phase Q$_1$ & IF phase Q$_1$ & \hspace{2em} \Qcircuit @C=1em @R= 0.7em {
\lstick{\ket{\textrm{Q}_1}} & \gate{\sqrt{\textrm{X}}} &  \multigate{1}{\textrm{CZ}^n} & \gate{\pm \sqrt{\textrm{Y}}} & \multigate{1}{\textrm{PSB}} & \qw \\
\lstick{\ket{\textrm{Q}_2}} & \qw & \ghost{\textrm{CZ}^n}  & \qw & \ghost{\textrm{PSB}} & \qw } \\ 

   \\ \hline\\
   Phase Q$_2$ & IF phase Q$_2$ & \hspace{2em} \Qcircuit @C=1em @R= 0.7em {
\lstick{\ket{\textrm{Q}_1}} & \qw &  \multigate{1}{\textrm{CZ}^n} & \qw & \multigate{1}{\textrm{PSB}} & \qw \\
\lstick{\ket{\textrm{Q}_2}} & \gate{\sqrt{\textrm{X}}} & \ghost{\textrm{CZ}^n} & \gate{\pm \sqrt{\textrm{Y}}} & \ghost{\textrm{PSB}} & \qw } \\ 
\\  \hline\\
     Readout point &  Detuning voltage & Charge anticrossing sweep \\ 
\end{tabular}
\end{ruledtabular}

\end{center}
\end{table*}

\begin{figure*}
    \centering  
    \includegraphics[width=2\columnwidth]{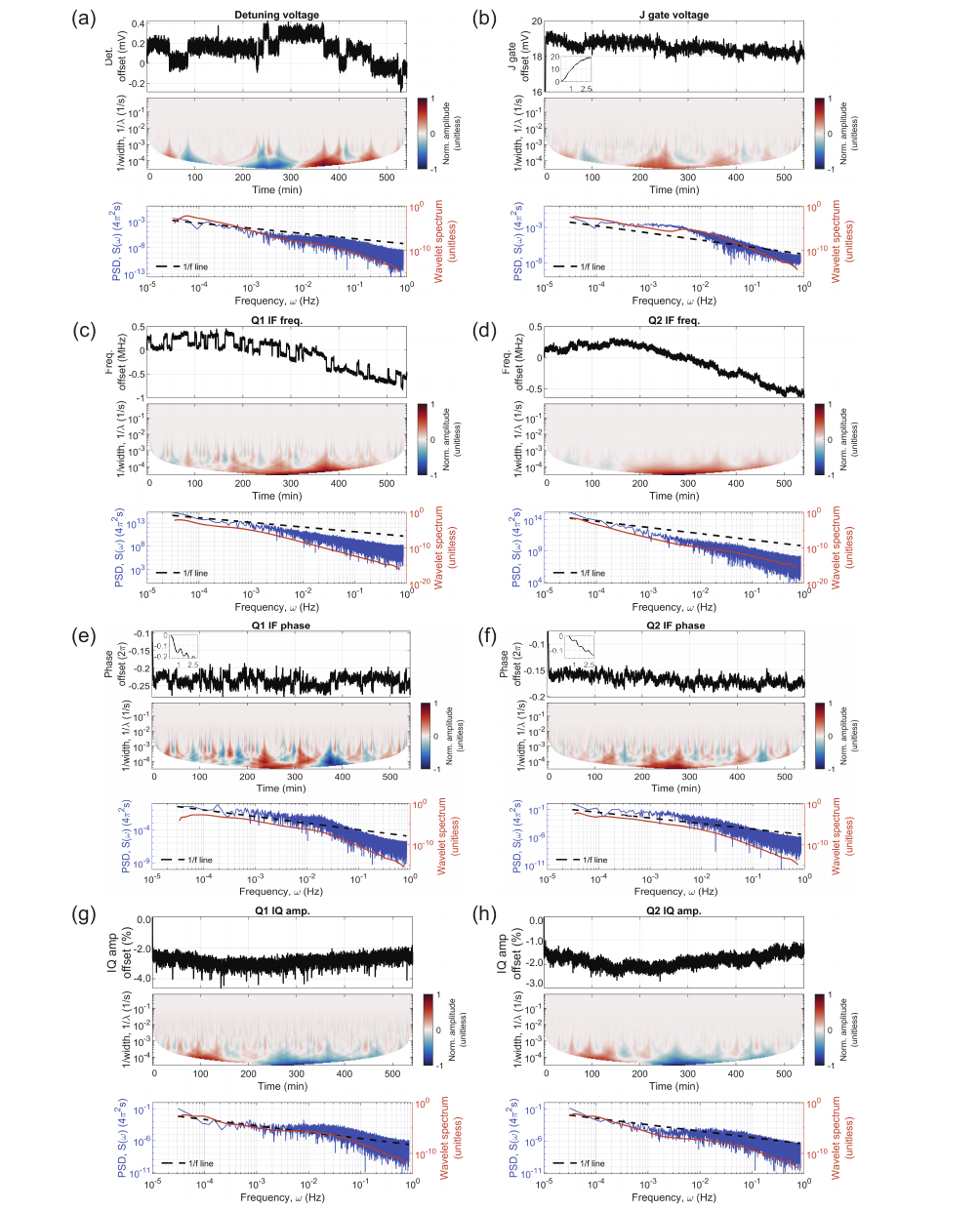}
    \caption{Overview of the data from the different feedback protocols as described in Tab.~\ref{tab:feedback} and detailed in the main text. Given for each protocol are raw feedback data, wavelet transformations $W(\lambda, \tau)$ normalised to the largest value of the wavelet transform, and Fourier power spectral densities and wavelet spectra. (a) Detuning voltage, (b) J gate voltage, (c) \& (d) frequency feedback for Q$_1$ \& Q$_2$, respectively, (e) \& (f) phase feedback for Q$_1$ \& Q$_2$, respectively, (g) \& (h) IQ amplitude feedback for Q$_1$ \& Q$_2$, respectively.  Insets in (b), (e), and (f) detail the parameter correction for the first 2.5 min, highlighting the feedback effectiveness against coarse calibration errors.}
    \label{fig:tlf}
\end{figure*}

We analyse experimental qubit parameter drift data obtained from the feedback protocols summarised in Table~\ref{tab:feedback}. We use small quantum circuits optimised for efficient execution on FPGA hardware to track and correct individual qubit parameters. First, we execute the circuit and measure the spin probabilities to capture the parameter deviation from its ideal value using a small number of shots (typically $\approx 20$ shots). Next, we update the control parameter with a correction proportional to the deviation. We refer to the control parameter corrections as the feedback data from now on. Each feedback data set is normalised with its initial value. We collect feedback data during 542 minutes by applying all feedback protocols sequentially, while keeping qubits idling between feedback protocol runs. Each feedback protocol is executed every 0.6 seconds. Figure~\ref{fig:tlf} shows an overview of the collected data over this approximately 10 hour experiment.
Feedback protocols focus on the following parameters: (1,3)-(0,4) charge transition detuning point, Q$_{1,2}$ Larmor and Rabi frequencies, exchange-gate voltage level, and Q$_{1,2}$ qubit phases for implementation of the two-qubit controlled Z (CZ) gate \cite{Vandersypen2004NMRComputation,Tanttu2023StabilitySilicon}. We will discuss details on each protocol next. 

Initialisation, readout, and qubit control voltage operation points can shift under influence of charge noise, impacting operation fidelities \cite{Tanttu2023StabilitySilicon}. The detuning feedback protocol measures the position of the (1,3)-(0,4) inter-dot charge transition and applies a correction to the detuning, $\epsilon \equiv V_{P1} - V_{P2}$, to all voltage points such that their position relative to the anti-crossing stays constant. We determine the (1,3)-(0,4) inter-dot charge transition position by sweeping $\epsilon$ over a range encompassing the charge transition and simultaneously integrate the charge sensor current. We chose an initial value for the integrated sensor current such that the charge transition is positioned in the middle of the chosen detuning voltage sweep window. A movement of the charge transition along the detuning axis is captured by a change in the integrated charge sensor current. Finally, the charge transition movement is cancelled by applying a detuning voltage correction proportional to the difference of the initial and measured sensor current value. The detuning feedback data is shown in Fig.~\ref{fig:tlf}(a).

The Larmor frequency feedback protocol is based on a modified Ramsey sequence. First, we apply an $\sqrt{\textrm{X}}$ gate to the target qubit and leave it idling for a time $t_{\text{wait}}= 100$ ns. Next, a $\pm \sqrt{\textrm{Y}}$ gate projects the qubit state on the $\pm y$-axis, as shown in Table~\ref{tab:feedback}. In the ideal case of no frequency detuning, the two projections both return spin flip probabilities of 0.5. Finally, we update the target qubit intermediate frequency (IF) stored in the FPGA unit using a proportional correction. This feedback data is plotted in Fig.~\ref{fig:tlf}(b,f).

Qubit Rabi frequencies are stabilised by correcting the mw burst amplitude using the FPGA output and IQ control. Correcting mw burst time is another option, but it is limited in the time domain by the FPGA waveform resolution of 4 ns and therefore less accurate \cite{Gilbert2023On-demandQubits}. An initial $\pi/2$ mw pulse time is calibrated for the target qubit by measuring a standard Rabi chevron and fitting the centre frequency. Next, the protocol measures spin flip proportions after applying $n$ and  $n + 2$ $\pi/2$ $X$ gates on the target qubit, where $n$ is an odd integer. In the ideal case both sequences give 0.5 spin flip proportions.  The under or over rotation is corrected by adjusting the target qubit IQ output power proportional to the measured outcome difference. The protocol circuit is shown in Table \ref{tab:feedback} and the Rabi frequency feedback data is plotted in Fig.~\ref{fig:tlf}(d,h).

Exchange interaction is controlled by pulsing the voltage on the J-gate \cite{Tanttu2023StabilitySilicon}. To achieve a constant exchange interaction during the two-qubit gate we apply a correction to the J-gate pulse voltage. Initially, the J-gate pulse duration and voltage level is calibrated such that it performs a controlled-Z (CZ) gate \cite{Vandersypen2004NMRComputation, Tanttu2023StabilitySilicon}. The feedback protocol then involves initialising Q$_{1}$ along the $y$-axis and Q$_{2}$ on the $z$ axis, repeated pulsing of the J-gate for the pre-calibrated CZ-gate time, typically $\approx 200$ ns, for $n = 7$ times, and finally projecting the target state onto the $\pm z$-axes for measurement. The circuit is shown in Table \ref{tab:feedback}. An ideal case results in flip probabilities of 0.5 and the difference between the two projections reflect under- or over-rotation in the CZ gate. An offset voltage to the J-level pulse, proportional to the measured difference, correct for the under- or over-rotation. The J-gate voltage level feedback data is shown in Fig.~\ref{fig:tlf}(e).   

Feedback for the individual qubit 1(2) mw phases starts by initialising qubit 1(2) on the $y$-axis and qubit 2(1) on the $-z$ axis. Next, we apply an exchange interaction pulse for a pre-calibrated CZ-gate time $\approx 200$ ns, and finally project the qubit 1(2) state onto the $\pm z$-axes for measurement, shown in Table \ref{tab:feedback}. An ideal case results in spin flip probabilities of 0.5. Finally, the difference between the two projections is used for a proportional correction to the qubit 1(2) mw phase. This feedback data is plotted in Fig.~\ref{fig:tlf}(c,g).
\\ \\
While feedback data is often secondary output of an experiment, it holds information about the fluctuations of qubit parameters and the specifics of the feedback implementation~\cite{Vepsalainen2022ImprovingFeedback}. Wavelet analysis allows us to analyse a data set $x(t)$ for signals of certain periodicity, referred to as the wavelet width or resolution $\lambda$, occurring at certain time(s) in the experiment, referred to as time translations or $\tau$. This is achieved by performing the continuous wavelet transformation, described by the equation
\begin{equation}
    W(\lambda, \tau) = \int_{-\infty}^{\infty} \psi_{\lambda, \tau}^*(t) x(t) dt,
    \label{eq:tranformation}
\end{equation}
where $W(\lambda, \tau)$ are the wavelet width and time translation coefficients, $\psi_{\lambda, \tau}$ the chosen wavelet function, and $t$ time. The results of the transformation can be visualised in two dimensions, shown in Fig.~\ref{fig:tlf}. This method is particularly useful for detecting signals that exhibit non-stationary properties, such as changes in frequency or amplitude over time~\cite{Seedhouse2023Wavelet-basedSystems, Prance2015IdentifyingDetection}. A useful property of wavelet analysis is the wavelet variance spectrum, defined as the variance squared for each $\lambda$ in $W(\lambda, \tau)$, which can be used to describe the spectral information about a data set, similar to that of the PSD in Fourier analysis.  


Fig.~\ref{fig:tlf} shows the gathered feedback data, $W(\lambda, \tau)$ (blue and red colour map), and the PSD and wavelet spectrum (blue and red line plots, respectively) for all feedback protocols shown in Tab.~\ref{tab:feedback}. The wavelet spectrum is defined as $\sigma^4$, where $\sigma^2$ is the variance of the wavelet transformation at each $\lambda$~\cite{Percival2000WaveletAnalysis}. The figure shows the wavelet spectrum normalised to the sum of the spectrum. The wavelet spectrum and the PSD are qualitatively similar across most of the spectrum, however, small variations can be observed in the spectra as the cone of influence~\cite{Percival2000WaveletAnalysis} (where boundary effects compromise the wavelet transformation) starts to significantly affect the wavelet transformation at lower values of $1/\lambda$ (below $1/\lambda \approx 10^4$ 1/s). Further analysis of the data is performed in Ref.~\cite{Seedhouse2023Wavelet-basedSystems}. In this Letter, we focus on the insights derived from the Haar wavelet transformation~\cite{Haar1911ZurFunktionensysteme}.


In this section we focus on the different features present in the data sets, and how they appear in the wavelet transformation. The feedback data in Fig.~\ref{fig:tlf} shows several features which we categorise as: 1) discrete jumps, 2) slow drifts of low amplitude, and 3) small time-scale jumps. 

The raw data shown in Fig~\ref{fig:tlf}(a), (b) and (c) features discrete jumps. The detuning voltage, Fig~\ref{fig:tlf}(a), shows a negative and positive shift at times $\approx$ 50 min and $\approx$ 90 min, respectively. To interpret these jumps in $W(\lambda, \tau)$ we recall that the wavelet transformation multiplies a Haar wavelet with the data at a particular $\lambda$ and $\tau$, Eq.~\ref{eq:tranformation}. If the wavelet shape matches a feature in the data, $W(\lambda, \tau)$ is amplified, i.e. a good match between a data feature and the wavelet is shown as a high \textit{positive} amplitude in $W(\lambda, \tau)$. A data feature matching the inverse of the wavelet results in a high \textit{negative} amplitude. The widths of the peaks at $t \approx$ 50 min and $t \approx$ 90 min increase as $1/\lambda$ decreases until the width of the features starts overlapping. When high and low amplitude features meet they cancel to zero, shown as off-white in Fig.~\ref{fig:tlf}. Features with the same sign join and amplify $W(\lambda, \tau)$. Examples in the detuning voltage $W(\lambda, \tau)$ shown in Fig~\ref{fig:tlf}(a) occur at $t \approx$ 375 and $t \approx$ 400 minutes, where two successive discrete jumps occur with the same sign, indicating that multiple two-level systems are involved. This is confirmed by the number of discrete levels seen in the detuning data which has four levels at detuning offset values of approximately 0.3 mV, 0.2 mV, 0 mV, and -0.1 mV. In comparison, data for Q$_{1}$ IF frequency and IF phase feedback displayed in Fig~\ref{fig:tlf}(c) and (e) show fewer consecutive same-sign fluctuations, resulting in pattern of peaks with alternating sign in the wavelet transform at around $1/\lambda \approx 10^{-3}$ 1/s. 
 
It is thought that 1/f noise originates from ensembles of trapped two-level fluctuators, presumably in the dielectric material~\cite{Hooge2003OnNoise}. In this model, a single fluctuator contributes to the overall frequency spectrum following a Lorentzian distribution, manifesting itself in the Fourier PSD as a plateau followed by a 1/f$^2$ trend starting at the fluctuator's characteristic frequency. Fourier PSD shown in Fig.~\ref{fig:tlf}(c), (d) and (h) deviate from the 1/f trend and exhibit a bump at $\omega \approx 3\times10^{-2}$ Hz, indicating a single, strongly coupled two-level fluctuator, or a more complex interplay of single and ensembles of two-level fluctuators\cite{Elsayed2022LowManufacturing}. While both the wavelet spectrum and Fourier PSD provide information on the characteristic frequency of the fluctuator, $W(\lambda, \tau)$ reveals more information of the dynamics of a fluctuator event in the time-domain. 
 
Data presented in Fig~\ref{fig:tlf}(c) and (d) shows long timescale drifts which manifest in $W(\lambda, \tau)$ as high amplitude features at small $1/\lambda$ values. For unidirectional drifts, $W(\lambda, \tau)$ has a constant sign at smaller $1/\lambda$ values. An example can be seen in Fig.~\ref{fig:tlf}(d) where the main amplitude in $W(\lambda, \tau)$ remains positive below $1/\lambda=10^{-4}$ 1/s. For multi-directional drifts, such as the one seen in Fig.~\ref{fig:tlf}(g) and (h), $W(\lambda, \tau)$ shows both positive and negative amplitudes. 

Persistent in all the feedback data are small amplitude, fast time-scale jumps. For example, the fast time-scale jumps in Fig.~\ref{fig:tlf}(a) have a standard deviation of approximately 0.05 mV, compared to the 0.2 mV standard deviation of the multi-level fluctuations discussed previously. The jumps resemble 1/f noise \cite{Struck2020, Chan2018AssessmentSpectroscopy} presenting itself as high frequency jumps occurring about the moving mean of the data and at smaller amplitudes than other features. The wavelet transformation shows these jumps as features at high $1/\lambda$ with close to zero normalised amplitude. The PSD and wavelet spectra further confirm the presence of 1/frequency noise \cite{Elsayed2022LowManufacturing}.

Another advantage of wavelets is their ability to compare signals at particular time-scales, aiding the characterisation of noise in the device. For example, similar features appear in $W(\lambda, \tau)$ in Fig~\ref{fig:tlf}(b) and (e) at $1/\lambda \approx 5\times10^{-4}$ 1/s and below. These insights help understanding important noise correlations within the system, a more rigorous correlation wavelet-based analysis is detailed in Ref.~\cite{Seedhouse2023Wavelet-basedSystems}. \\ 



In conclusion, we present several FPGA-optimised feedback protocols to stabilise single- and two-qubit parameters for electron spin qubits formed in silicon quantum dots over a multi-hour timescale. We analyse the feedback data with a Haar wavelet transform, and compare the results with the Fourier PSD. Wavelet analysis shows evidence of several different noise signals, namely 1) discrete jumps, 2) slow drifts of low amplitude, and 3) small time-scale jumps. Further work will include analyses using different wavelet bases, including complex wavelets, and investigate correlations between the different qubit parameters. Such detailed breakdown of the noise signals can help to improve qubit fabrication processes and implement cost effective and scalable feedback protocols.
These results demonstrate that FPGA-based qubit parameter tracking protocols and wavelet analyses are valuable tools to optimise qubit performance and characterise them against various noise sources.

\section*{Acknowledgements}
We acknowledge support from the Australian Research Council (FL190100167 and CE170100012), the US Army Research Office (W911NF-23-10092), and the NSW Node of the Australian National Fabrication Facility. The views and conclusions contained in this document are those of the authors and should not be interpreted as representing the official policies, either expressed or implied, of the Army Research Office or the US Government. The US Government is authorised to reproduce and distribute reprints for Government purposes notwithstanding any copyright notation herein. A.E.S., S.S., and J.Y.H. acknowledge support from the Sydney Quantum Academy. 

\section*{Data availability}
The data that support the findings of this study are available from the corresponding author upon reasonable request.

\section*{Author contributions}
W.H.L. and F.E.H. fabricated the devices, with A.S.D.’s supervision, on isotopically enriched $^{28}$Si wafers supplied by K.M.I.(800ppm). N.D.S., W.G., S.S., T.T., J.Y.H., and W.H.L. did the experiments, coding and initial analysis, with A.L., A.S., C.H.Y., and A.S.D.’s supervision. N.D.S. and A.E.S. did the feedback analysis. N.D.S. and A.E.S. wrote the manuscript, with the input from all authors.

\bibliography{references_nard}

\end{document}